A decade of molecular cell atlases

Lior Pachter[*]
Division of Biology and Biological Engineering and Department of Computer Science
California Institute of Technology

*Correspondence: lpachter@caltech.edu

The recent opinion article "A decade of molecular cell atlases" by Stephen Quake [1] narrates the incredible single-cell genomics technology advances that have taken place over the last decade, and how they have translated to increasingly resolved cell atlases. However the sequence of events described is inaccurate and contains several omissions and errors.

Quake writes that "The year 2017 marked the release of our *Tabula Muris* data set and preprint" for which he cites the 2nd version of a preprint on the bioRxiv posted on March 29, 2018 [2]. While it is true that the preprint was first posted on December 20, 2017 [3], at that time the data set was not released. While the gene counts and cell metadata were made available, these resources were produced with analyses of the unreleased data set using a complex data processing workflow that included setting 16 parameters for alignment of the reads followed by a nine step clustering procedure to label cell types (see Methods of [3]). The data set was only released on March 19, 2018 (Gene Expression Omnibus accession GSE109774), after which it was published with [2]. Until the data set was released, it was not possible to verify or reproduce results from a paper, nor to build on the *Tabula Muris* by uniformly processing the data together with other data sets for joint analysis.

The error in describing the date when the *Tabula Muris* data was shared is significant in light of Quake's claim that the *Tabula Muris* was "the first mammalian whole-organism cell atlas". Quake describes another single-cell RNA-seq based mouse cell atlas [4] as "further work", despite the fact that both the paper and its associated data set were publicly available prior to the *Tabula Muris*. In fact, the data set of [4] is analyzed in [2] with the authors concluding that "independent datasets generated from various atlases that are beginning to arise can be combined and collectively analyzed…". Thus, it was [2], and its eventual journal publication on October 3, 2018 [5] that was "further work" following [4], not the other way around.

Quake further attempts to distinguish [2] by noting that it resulted from an "important decision at CZIBiohub…to pursue whole-organism cell atlases rather than individual organs separately", in contrast to other projects that he characterizes as "representing a distinct strategy of compiling cell atlases one tissue at a time". As examples of the latter he lists [6], [7] and [8], but [6] covers 60 human tissue types, while both [7] and [8] each mapped cell types in 15 human organs. Quake also fails to mention that the human cell atlases [6], [7] and [8] were published before the *Tabula sapiens* preprint [9].

In reviewing the technology developments that led to high-throughput single-cell RNA-seq, and the eventual curation of mouse and human cell atlases, Quake centers his own work while omitting several important advances, such as [10] which pertains to the key step of of barcoding distinct molecules (unique molecular identifiers) to facilitate molecule counting in the face of PCR duplicates. Similarly, Quake fails to acknowledge the central role of Aviv Regev and Sarah Teichmann in developing human cell atlases. They were early champions of a collaborative human cell atlas project and have co-chaired the organizing committee to facilitate its success from the outset. Teichmann co-founded the Wellcome Trust Sanger Institute Single Cell Genomics Centre in 2013, and by 2015 had been awarded the EMBO Gold Medal in part for her contributions to, and vision for, single-cell transcriptomics. Regev pioneered many of the single-cell RNA-seq technology developments that enabled single-cell genomics, such as a landmark single-cell studies of immune cells in 2013 [11]. By the time of the inaugural Human Cell Atlas meeting in London in 2016, Regev had been widely publicizing a vision for a "periodic table of cells" and Teichmann, who had just published the important review [12], had joined forces with her to develop a joint vision to accomplish the task. Their contribution to the human cell atlas project went far beyond just asking "whether various [cell atlas] efforts should be merged into an international collaboration", which is what Quake credits them for.

There are also a few minor errors in the paper. Quake writes that "These [microfluidics automation technologies] were eventually commercialized by a company I founded called Fluidigm..". In fact, Quake did not found Fluidigm by himself; the company was co-founded with Gajus Worthington. Miriam Merad's name is incorrectly spelled as "Meriam" and Christophe Benoist's name is incorrectly spelled as "Benoiste". A recurring typo is the misspelling of Sarah Teichmann's name. It is incorrectly spelled "Teichman" three times throughout the manuscript.